\documentstyle[preprint,aps,prl,amssymb]{revtex}

\begin{document}

\tightenlines

\title{ Vacuum structure of  $\IC {\rm P}^{\rm N}$ sigma models at 
$\theta=\pi$}

\author{\bf {M. Asorey} and F. Falceto}
\address{Departamento de
F\'{\i}sica Te\'orica. Facultad de Ciencias\\
Universidad de Zaragoza.  50009 Zaragoza. Spain}

\def\CAG{{\cal A/\cal G}} \def\CO{{\cal O}} 
\def\CA{{\cal A}} \def\CC{{\cal C}} \def\CF{{\cal F}} \def\CG{{\cal
G}} \def\CL{{\cal L}} \def\CH{{\cal H}} \def\CI{{\cal I}}
\def\CU{{\cal U}} \def\CB{{\cal B}} \def\CR{{\cal R}} \def\CD{{\cal
D}} \def\CT{{\cal T}} \def\CK{{\cal K}}
\def\e#1{{\rm e}^{^{\textstyle#1}}}
\def\grad#1{\,\nabla\!_{{#1}}\,}
\def\gradgrad#1#2{\,\nabla\!_{{#1}}\nabla\!_{{#2}}\,}
\def\ph{\varphi}
\def\semi{;}
\def\psibar{\overline\psi}
\def\om#1#2{\omega^{#1}{}_{#2}}
\def\vev#1{\langle #1 \rangle}
\def\lform{\hbox{$\sqcup$}\llap{\hbox{$\sqcap$}}}
\def\darr#1{\raise1.5ex\hbox{$\leftrightarrow$}\mkern-16.5mu #1}
\def\lie{\hbox{\it\$}} 
\def\ha{{1\over2}}
\def\half{{\textstyle{1\over2}}} 
\def\roughly#1{\raise.3ex\hbox{$#1$\kern-.75em\lower1ex\hbox{$\sim$}}}
\def\inbar{\,\vrule height1.5ex width.4pt depth0pt}
\def\minbar{\,\vrule height1.0ex width.4pt depth0pt}
\def\IB{\relax{\rm I\kern-.18em B}}
\def\IC{\relax\hbox{$\inbar\kern-.3em{\rm C}$}} \def\ID{\relax{\rm
I\kern-.18em D}}  
\def\IE{\relax{\rm I\kern-.18em E}}
\def\IF{\relax{\rm I\kern-.18em F}}
\def\IG{\relax\hbox{$\inbar\kern-.3em{\rm G}$}}
\def\IH{{\Bbb H}}
\def\II{\relax{\rm I\kern-.18em I}}
\def\IK{\relax{\rm I\kern-.18em K}}
\def\IL{\relax{\rm I\kern-.18em L}}
\def\IM{\relax{\rm I\kern-.18em M}}
\def\IN{\relax{\rm I\kern-.18em N}}
\def\IO{\relax\hbox{$\inbar\kern-.3em{\rm O}$}}
\def\IP{\relax{\rm I\kern-.18em P}}
\def\IQ{\relax\hbox{$\inbar\kern-.3em{\rm Q}$}}
\def\IR{{\Bbb R}}
\font\cmss=cmss10 \font\cmsss=cmss10 at 10truept
\def\IZ{\relax\ifmmode\mathchoice
{\hbox{\cmss Z\kern-.4em Z}}{\hbox{\cmss Z\kern-.4em Z}}
{\lower.9pt\hbox{\cmsss Z\kern-.36em Z}}
{\lower1.2pt\hbox{\cmsss Z\kern-.36em Z}}\else{\cmss Z\kern-.4em
Z}\fi} \def\ZZ{{\Bbb Z}}
\def\IGa{\relax\hbox{${\rm I}\kern-.18em\Gamma$}}
\def\IPi{\relax\hbox{${\rm I}\kern-.18em\Pi$}}
\def\ITh{\relax\hbox{$\inbar\kern-.3em\Theta$}}
\def\IOm{\relax\hbox{$\inbar\kern-3.00pt\Omega$}}
\def\semi{;\hfil\break}
\def\CA{{\cal A}}\def\CCA{$\CA$}
\def\CC{{\cal C}}\def\CCC{$\CC$}
\def\CT{{\cal T}}\def\CCT{$\CT$}
\def\CQ{{\cal Q}}
\def\CS{{\cal S}}
\def\CP{{\cal P}}\def\CCP{$\cal P$}
\def\CO{{\cal O}}\def\CCO{$\CO$}
\def\CM{{\cal M}}\def\CCM{$\CM$}
\def\CMH{\widehat\CM}\def\CCMH{$\CMH$}
\def\CMB{\overline\CM}\def\CCMB{$\CMB$}
\def\CH{{\cal H}}\def\CCH{$\CH$}
\def\CL{{\cal L}}\def\CCL{$\CL$}
\def\CS{{\cal S}}\def\CCS{$\CS$}
\def\CX{{\cal X}}
\def\CE{{\cal E}}\def\CCE{$\CE$}
\def\CV{{\cal V}}\def\CCV{$\CV$}
\def\CU{{\cal U}}\def\CCU{$\CU$}
\def\CF{{\cal F}}\def\CCF{$\CF$}
\def\CG{{\cal G}}\def\CCG{$\CG$}
\def\CN{{\cal N}}\def\CCN{$\CN$}
\def\CD{{\cal D}}\def\CCD{$\CD$}
\def\CZ{{\cal Z}}\def\CCZ{$\CZ$}
\def\cte{\rm cte}
\def\cs{W(z)}
\def\css{S_{\rm CS}}
\def\cst{Chern-Simons theory}
\def\csts{Chern-Simons theories}
\def\Tr{{\rm Tr}}
\def\tr{{\rm tr}}
\def\Nabla{{\Bbb D}}
\def\sph{{z_{\rm sph}}}
\def\asph{{\widetilde{z}_{\rm sph}}}
\def\vac{{z_{\rm vac}}}
\def\be{\begin{eqnarray}}
\def\ee{\end{eqnarray}}
\def\th{$\theta$}
\def\cp{{$\IC {\rm P}^{\rm N}$}} 
\def\cpp{{{\rm S}^{\rm 2N-1}}}
\def\cpa{{$\IC {\rm P}^{\rm 1}$}}
\def\sm {sigma model}
\maketitle
\begin{abstract}

	We show that parity symmetry is not spontaneously broken in
the \cp\ sigma model for any value of N when the coefficient of
the  $\theta$--term  becomes $\theta=\pi$ (mod. $2\pi$).
The result follows from a non-perturbative analysis of the nodal structure 
of the vacuum functional  $\psi_0(z)$. The dynamical role of  
 sphalerons turns out to be
 very important for the argument. The result introduces severe
constraints on the possible critical behavior of the models at 
$\theta=\pi$ (mod. $2\pi$).

\end{abstract}
\hyphenation{}
\pacs{PACS: 11.30.Er, 11.15.Tk, 75.10 Jm
{\tt$\backslash$\string pacs\{\}} should always be input,
even if empty.}
 \narrowtext\noindent
 
 \narrowtext\noindent

The effect of a CP-violating $\theta$-term interaction  on the 
infrared behavior of 
quantum field theories generates  
 substantial changes in the low energy
spectrum   
\cite{callan-jackiw}. 
For this  reason, phenomelogical requirements impose severe bounds on the
actual value of the $\theta$-parameter.
  The analysis of the 
response of the systems to such a topological perturbation
is, however, a very rich source of information on  the
vacuum structure of the quantum systems at $\theta=0$ 
\cite{www}. This structure is very hard to
analyze by direct methods because it entirely dwells
on the strong
coupling regime and involves deep non-perturbative behaviors.
The most
relevant changes appears at $\theta=\pi ({\rm mod }\ 2 \pi) $. In this
case the classical lagrangian is also CP invariant as in absence of the
$\theta$--term, $\theta=0 ({\rm mod }\ 2 \pi)$, 
 but the behavior of the system  is completely different in
these two regimes.

 QCD, the archetype of such systems,  is still  
 inaccessible for 
analytic studies of its prominent physical effects: confinement and
chiral symmetry breaking.  \cp\  sigma models
share many similar
features like
dynamical mass generation, confinement and asymptotic freedom, but
they are simpler to analyze. 
In particular for some of those models there is  exact analytic
information on their quantum spectrum. 
It is, therefore, interesting to
analyze the $\theta$--vacuum effects in those systems in order to
gain some insights into the similar effects in QCD$_{3+1}$.

The simplest  model, \cpa, is integrable for $\theta=0$ and
$\theta=\pi$, the only two values of $\theta$ for which 
the system is classically CP invariant. However, the behavior of the
system is very different in those cases. At $\theta=0$ the system
is confining and exhibits  a mass gap; at $\theta=\pi$ the model is
massless and its critical exponents are those of SU(2)
Wess-Zumino-Witten conformal invariant model at level 1 \cite{zamm,zam}. 
In both cases the CP symmetry is not spontaneously broken.
In the second case this can be rigorously shown in a
discrete regularization of the model where it 
turns out to be 
equivalent  to a spin $\half$ chain by Haldane transformation
\cite{Hal};
for those chains the
Lieb-Schulz-Mattis theorem \cite{lms} establishes that there
are only two possibilities: either parity is spontaneously broken
and there is a mass gap or the theory is gapless
and parity is not spontaneously broken. Since it is known that the mass gap is
zero for $\theta=\pi$ \cite{zamm}--\cite{Hal}\cite{sh}, then parity is not
spontaneously broken.

For higher values of N the integrability of the 
\cp\ model is lost and the  only  
non-perturbative  information comes  either from large N  \cite{adda}
and strong coupling  expansions \cite{sei}
or   Monte Carlo numerical simulations 
\cite{bhanot}--\cite{plekss}.
  Strong coupling expansions indicate at leading
order the existence of first order phase transition 
at $\theta=\pi$ accompanied by a spontaneous
symmetry breaking of parity.  
Large N expansions show that there is a non-vanishing mass gap at
$\theta=\pi$ and at leading order also foresee a similar critical behavior.
Although large N and strong coupling
limits do not commute \cite{rss} the  scenarios emerging from
both approximations are compatible. Moreover, it has been conjectured
that this kind of critical behavior holds for all models with $n>1$.
Numerical Monte Carlo calculations  
pointed out the appearance of another unexpected first
order transition for lower values of $\theta<\pi$ in the CP$^3$ model 
\cite{desy}. 
However, the interpretation of the calculations is not completely
clear because of the existence of large errors in the region near 
$\theta=\pi$ which make Monte Carlo results not very reliable 
\cite{plekss}.

The aim of this letter is to shed some new light into the
 behavior of \cp\ models at $\theta=\pi$ by a novel method
which was successfully introduced for a similar problem in QCD.
The method is based on the analysis of the nodal structure of
the vacuum functional of the theory. In QCD$_{3+1}$  and QED$_{1+1}$
(with an external
gauge invariant perturbation)
those nodes appear for $\theta=\pi$  at some classical gauge
configurations which include sphalerons \cite{qcd}\cite{ae}.  In both
cases  parity is not spontaneously broken \cite{qcd}\cite{ae}. In
QCD$_{2+1}$ with a Chern-Simons massive perturbation nodes also appear
at configurations with magnetic charge \cite{nod}. It has been  
conjectured that those nodal configurations play a  relevant role
for the  confinement mechanism of those theories in absence of the
topological perturbations. The generalization of those methods for
the \cp models is quite straightforward  but requires a
detailed analysis.

The basic $\sigma$--field variable   is  a
 complex field 
$z(x)$ with values in $\IC^{\rm N+1}$ constrained by the condition $z^\dagger z=1$.
In the temporal gauge, $A_0= 
{i\over 2} [z^\dagger\partial_0 z
-   (\partial_0 z^\dagger) z]=0$, the dynamics of the theory is governed  
by the quantum Hamiltonian, which in  Schr\"odinger representation reads 
\be
 \IH_\theta &=&{ g^2 \over 2}\int dx\    \left( 
{\delta\over
\delta z^\dagger(x)} - {i\theta\over 
2 \pi}\partial  z\right) \left({\delta\over\delta  {z}(x)} + 
{i\theta\over 
2\pi}{\partial  z}^\dagger\right) \cr
&&
 + {1\over 2 g^2}\int dx  (D z)^\dagger D z ,
\label{hamq}
\ee
\noindent
 where $\partial =\partial_x$ and $D=\partial-[z^\dagger\partial z
-   (\partial z^\dagger) z]/2$.
 Physical states must satisfy the Gauss law constraint
\be
 \left[ z{\delta\over \delta z}-  z^\dagger {\delta\over \delta z^\dagger}
\right]
\psi_{\rm
ph}(z)=0.\label{gaussss} \ee
This means that 
$\psi_{\rm ph}(z^{\Phi})=\psi_{\rm ph}(z)$ for any U(1) gauge 
transformation  
$z^\Phi(x)=z(x) \Phi(x)$, with $\Phi={\rm e}^{i\varphi(x)}$,
i.e. physical states must be invariant 
under  U(1) gauge transformations. 
Therefore, they can be identified
with functionals on the quotient space  
$\CP=\Sigma/\CG$ of
 the space $\Sigma$ of 1-dimensional $\sigma$-fields 
 by the group  $\CG$ of U(1) gauge transformations.

The quantum theory presents ultraviolet divergences which
require renormalization of the $\sigma$-fields  $z$ and the
coupling constant g. It is, therefore, necessary to introduce an
ultraviolet  regularization in the Hamiltonian (\ref{hamq}). 
We shall consider the regularization
 \cite{gav}

\be
\IH_\theta^{\rm reg}&=&
{ g^2 \over 2}\int dx\     \left({\delta\over\delta {z}^\dagger(x)} -
 {i\theta\over 
2\pi}{\partial z}\right)
\cr && \left(I-{D^2\over\Lambda^2}\right)^{-n} \left( 
{\delta\over
\delta z(x)} + {i\theta\over 
2 \pi}\partial z^\dagger\right) 
\cr &&  + {1\over 2 g^2}\int dx  (D z)^\dagger D z  ,
\ee
\noindent
which preserves most of the symmetries of the model and
is similar to that introduced for gauge theories in Ref. \cite{amm}. 
In both cases the
Schr\"odinger formalism of the
the quantum theory remains ultraviolet finite.

The non-trivial effect of the $\theta$--term is due to the
non-simply connected character of the orbit space $\pi_1(\CP)=\ZZ$
or what is equivalent the non-connected character of the group of
gauge transformations $\pi_0(\CG)=\ZZ$.   The regularized
Hamiltonian  can be written as
$$
{ \IH^{\rm reg}_\theta= {g^2\over 2} \int d x \   
\Nabla^\theta_{z^\dagger }
\left(I+{D^2\over\Lambda^2}\right)^{-n}\Nabla_\theta^{z}  +{1\over 2
g^2}\int d x  \ (D z)^\dagger D z ,} $$
%
where
\be
{\Nabla^\theta_z={\delta\over \delta z}+{i\theta\over 2 \pi}
\partial  z^\dagger \qquad
\Nabla^\theta_{z^\dagger}={\delta\over \delta
{z^\dagger}}-{i\theta\over 2 \pi} \partial  z,}\label{der}
\ee
is a functional covariant derivative with respect to the U(1)
ultra-gauge field defined over the space of sigma fields $\Sigma$ by
the ultra-gauge flat vector potential
\be
{\alpha^\theta_z={\theta\over
2\pi}\partial  z^\dagger\qquad 
\alpha^\theta_{z^\dagger}=-{\theta\over
2\pi}\partial  z.}\label{form} \ee
The projections of the action of the operators $D_{z^\dagger}$ and $D_{z}$ 
on gauge invariant
functionals become covariant derivatives over $\CP$ with respect 
the  non-trivial flat connection defined by the projection of
$\alpha^\theta$ on $\CP$.
Now since the space of gauge orbits of field configurations $\CP$
is not simply connected, the effect of the $\theta$--term
becomes non-trivial because it cannot be removed by any smooth
gauge transformation for any value of $\theta\neq 2\pi n$.
This happens because
the projection of the potential
$(1/\theta)\alpha^\theta$ is a generating form of the non-trivial
first cohomology group $H^1(\CP,\ZZ)=\ZZ$ of $\CP$. 
In this sense
the phenomenon is very similar to the Aharanov-Bohm effect\cite{amm}.

It is, however, possible to remove the $\theta$ dependence of the Hamiltonian 
by means of a singular gauge transformation
$
{\xi(z)={\rm e}^{-{i\theta\over 2\pi}W(z)}  \psi_{\rm ph}(z)},
$
with
\be
{W(z)=
{1\over 2\pi}\int  dx \   z^\dagger\partial  z .
}\label {hcs} 
\ee
The transformation yields a $\theta$--independent Hamiltonian    
$
{\widetilde{\IH}_\theta^{\rm reg}=
{\rm e}^{-{i\theta\over 2\pi}W(z)}
\IH_\theta^{\rm reg} {\rm e}^{{i\theta\over 2\pi}W(z)}=\IH_0^{\rm
reg}},$
but it does not preserves  Gauss law
because $W$ is not invariant under large gauge transformations $\Phi$ with
non-trivial winding number $\nu(\Phi)$,
$W(z^\Phi)=W(z)+ 2\pi \nu(\Phi)$. In other
words, $\xi$ is not globally defined on $\CP$. It is, however, possible
to gauge fix the
symmetry under global gauge transformations and find an open domain
$\Sigma_0=\{z\in \Sigma; -\pi
<W(z)< +\pi)\}$
in the space $\Sigma$ of $\sigma$ field configurations 
 such that any  other field configuration $z(x)$
is gauge equivalent to one which lies in the topological closure $\overline{\Sigma}_0$
of $\Sigma_0$. $\xi$ is uniquely defined in $\Sigma_0$ and only breaks
global gauge invariance
at the boundaries ${\partial_\pm \Sigma_0=\{ z\in \Sigma; W(z)= \pm \pi \}}$,
because
 for any pair of  fields
$z_-\in \partial_-\Sigma_0$ and $z_+\in
\partial_+ \Sigma_0$
 which are gauge equivalent
 $ z_+=z_-^\Phi$  by a gauge transformation with
winding number $\nu(\Phi)=1$, e.g.  $\Phi_1(x)=e^{2i\arctan (x/2a)}$.
 The $\theta$
dependence of the systems is now encoded by the non-trivial boundary conditions 
that
physical states have to verify  at the boundaries $\partial_-\Sigma_0$
and $\partial_+ \Sigma_0$  of  $\Sigma_0$,

In this sense the transformation  is trading the
$\theta$--dependence of the Hamiltonian by non-trivial boundary
conditions
on $\partial_+\Sigma_0$.
The projection of the boundaries $\partial_\pm\Sigma_0$  into $\CP$
define a codimension one submanifold $\CN$ 
which intersects any non-contractible
loop of $\CP$. This explains why within  this gauge fixing framework 
it is possible to remove the $\theta$--dependence of the system.

Classical constant vacua configurations $z_{\rm vac}(x)=z_0$ belong
to the domain $\Sigma_0$, and are gauge equivalent to any
 other vacua classical
 configuration  $ z'_{vac}=
 z_0 \Phi(x)$.
In the boundary of  $\Sigma_0$ there are
 sphalerons $\sph$ and anti-sphalerons $\asph$, which
 are quasi-stable
static  solutions of the classical motion equations with
only one unstable direction. They can exist only for finite volumes.
 The explicit expression for the first sphaleron on a finite circle $S^1$
 is given by the sigma field configuration  induced by the
2-dimensional  instanton on the circle  centered at the center of
the instanton \cite{golo} and with radius $a$ equal to the size of the
instanton $\rho$ \cite{vB} .
 In  stereographic projection of polar coordinates 
it reads
$
z_{\rm sph}(x)=({{{\bf u}+   {\bf v} e^{2i \arctan  (x/2a)} })/
\sqrt{2}} $
for any pair  ${\bf u}$ and ${\bf v}$ of
 orthogonal unit vectors of $\IC^{\rm N+1}$, i.e. 
${\bf u}^\dagger {\bf u}={\bf v}^\dagger
{\bf v}=1$ and ${\bf u}^\dagger {\bf v}=0$. 
The unstable mode corresponds to the radial
dilation generated by the flow
of the instanton. For infinite volume
those configurations  are not anymore 
  saddle points of the potential term
\be
V(z)={1\over 2 g^2} \int  dx(D
z)^\dagger D z. \label{pot}
\ee
However, the other relevant property of 
 sphalerons, 
$
W(\sph)=\pi
$
($\sph\in \partial_+\Sigma_0$), remains  independent of the   space volume
because the W(z)
functional is metric independent. In particular, this means that
 $\sph\in \partial_+\Sigma_0$. 
 The configuration generated by the anti-instanton flow
$
{\asph}{\ }(x)=({{{\bf u}+{\bf v} e^{-2i \arctan  (x/2a)}) }/
\sqrt{ 2}} 
$
in $\partial_-\Sigma_0$  also
exhibits similar properties.

There is another interesting feature of these two configurations.
 The U(N+1) transformation $S_{\bf u}^{\bf v}$ which rotates on the plane 
defined by ${\bf u}$ and ${\bf v}$ and interchanges
${\bf u}$ and ${\bf v}$, transforms the
anti-sphaleron $\asph$\ into a configuration which is gauge equivalent 
to the sphaleron, $S_{\bf u}^{\bf v}\asph={z}^{\Phi_1}_{\rm sph}$
by a gauge transformation
${\Phi_1(x)= e^{2i \arctan (x/2a)}}$
with winding number $\nu(\Phi_1)=1$. The non-trivial
boundary conditions  induced by the singular gauge 
transformation into physical states
imply that
${\xi(\sph)={\rm e}^{-{i\theta}}
\xi(S_{\bf u}^{\bf v}\asph)}$
for any sphaleron configuration.

The  theory is formally CP invariant only for  
$\theta=0$ and $\theta=\pi$ (mod $2\pi$). This can be seen from
the behavior of the boundary condition under CP which reverses
the sign of $\theta$ because the W(z) functional  is CP odd.
At $\theta=\pi$ the boundary condition, however, 
becomes an anti-periodic boundary condition,
${\xi(z_+)=- \xi(z_-)}$ which is CP invariant.
Because of  U(N+1) invariance the same property holds for the 
idempotent transformation $P_s=S_{\bf u}^{\bf v}$ P. 

The behavior of sphalerons and classical vacuum configurations under the
$P_s$ symmetry is rather different. The constant classical vacuum 
$z_{\rm vac}=z_0=\cte$ is
P invariant
($z_{\rm vac}^P=z_{\rm vac}$), but transforms under $P_s$ into
another classical vacuum configuration $ z^{P_s}_{\rm vac}=
S_{\bf u}^{\bf v} z_{\rm vac}$.  The sphaleron  is not parity invariant, 
but it is  
transformed 
by $P_s$ into a gauge
equivalent configuration, i.e.
$ z_{\rm sph}^{P_s}=S^{v}_u{\asph}=z_{\rm sph}^{\Phi_1}$, { with} $\nu(\Phi_1)=1$.
In fact, the whole  submanifold $\CN \in \CP$
is $P_s$ invariant, but  the sphaleron has an additional
peculiarity, it is quasi-invariant under this transformation.

In the full Hilbert space $\CH$ of all physical states 
which satisfy the anti-periodic boundary condition, there is always a
complete basis of stationary wave
functionals with a definite $P_s$ symmetry. If an energy level is not
degenerate the corresponding physical state $\xi(z)$
has to be $P_s$ even or $P_s$ odd. In the degenerate case, if $U(P_s)\xi(z)$
is not on the same ray that $\xi(z)$, the
stationary functionals $\xi_\pm(z)=
\xi(z)\pm U(P_s)\xi(z)$ are $P_s$ even/odd, respectively. If
$P_s$ is spontaneously broken the  quantum vacua $\xi_0(z)$ will not
have a definite $P_s$ parity ($\xi_0\neq\xi_\pm$) 
in the different physical phases and the  Hilbert space
will split into superselection sectors not connected by local
observables.

Now, it is easy to show that anti-periodic
boundary conditions imply the existence of nodes
in physical states with a definite $P_s$ parity. Actually, since
$
U(P_s)\xi_0( \sph)= \xi_0(\asph)
 = \xi_0(z^{\Phi_1}_{\rm sph }) =-\xi_0(\sph),\label{fpr}
$
if the vacuum state is $P_s$ even this is
possible only if $\xi_0$
vanishes for sphaleron configurations, $\xi_0(\sph)=0$.
 Odd functionals with respect to $P_s$--parity  
change sign from $z_0$ to $P_s z_0$, because $U(P_s)\xi_0(z_0)=
\xi(z^{P_s}_0)=-\xi(z_0)$.
Therefore, by continuity, they have to vanish  for some  
 constant classical vacua. If a quantum
vacuum is odd this means that it vanishes for all classical vacua because
of the U(N+1) invariance implied by the Coleman-Mermin-Wagner theorem.
Now the potential term of the
Hamiltonian $V(z)$ 
gives a finite positive contribution to the
energy of stationary states. 
The variation of $V$ along the trajectory of classical configurations 
defined by the flow of 
an instanton which interpolates between the sphaleron $\rho=a$ and
the classical vacua $\rho=0$ or $\rho=\infty$, given by 
$$V(z_\rho)={\pi\over 2 g^2}  {a\rho^2\over (a^2+\rho^2)^2}$$
indicates that the $P_s$ even ground states  which vanish at
sphalerons cannot have the same energy as  $P_s$ odd states which vanish at
classical vacuum configurations where the potential terms attains its minimal
value. Specially because the result holds for any value, weak or strong, of the
coupling constant $g^2$. Notice that
in the regularized theory   
there is no running of the coupling constant and, then,
the potential  term of the
Hamiltonian is not suppressed
in the infrared.

This feature implies that the quantum vacuum state $\psi_0(z)$ has to
be  even under $P_s$ parity, it has to vanish at sphaleron configurations,
($\psi_0(\sph)=0$), and the $P_s$ 
symmetry is not
spontaneously broken in the regularized theory. 

 On the other hand because of U(N+1) invariance   the vacuum functional $\xi_0$ 
can never be odd under $S_{\bf u}^{\bf v}$ parity. As a consequence 
$\xi_0$ has to be even under 
P  parity.  The
argument  for $P_s$ odd states applies also for the theory at $\theta=0$ and 
shows why parity is not spontaneously broken in that case too, where 
the quantum ground state is expected to have no nodes.

The vanishing of the quantum vacuum
functional $\psi_0$ for sphalerons, is basically based on 
two properties of those configurations: its 
  quasi-invariance  under $P_s$ parity
 and the special value that $W$ functional 
 reaches at them, $W(\sph)=\pi$. The arguments
  can be extended for any gauge field satisfying the same
properties.    The infinitesimal perturbations $z=\sph+\epsilon\tau
+\CO(\epsilon^2)$ of $\sph$
given by
$\tau(x)=S_{\bf u}^{\bf v}\omega(-x) + \omega\Phi_1^\ast(x)+i\sph(x)\varphi(x)$
preserve both properties, for any perturbation of the $\sigma$--fields
$\omega$ and any infinitesimal gauge transformation $\varphi$. 
Therefore they generate an infinite subspace of nodal configurations
for $\psi_0$ in $\partial_+\Sigma_0$. The same configurations are also nodes of
any higher energy stationary states with even $P_s$--parity $\psi_{\rm
even}(\sph)=0$.
 In a similar way it can be shown that the nodal configurations of 
$P_s$--parity odd
excited states contain  classical  vacuum configurations $\psi_{\rm
odd}(z_{\rm vac})=0$. The same results are obtained by
a path integral analysis along the lines developed in 
Ref. \cite{qcd}.

The existence of those nodes for any value of
the ultraviolet regulator $\Lambda$ and the coupling constant $g$,
implies that they also hold in the renormalized theory. The
absence of spontaneous breaking of the symmetry
also persists in the renormalized theory because the phenomenon 
is always related to
the infrared properties of the theory which are the same for the
regularized and renormalized theories.
Actually, the regularization of the kinetic term always enhances
the role of smooth configurations and then
the appearance of first order transitions, and since they do not
appear at $\theta=\pi$ in the regularized theory, the cannot reappear when 
the ultraviolet regulator is removed. For the \cpa\ model this is in perfect 
agreement with the exact results \cite{zamm}--\cite{sh}, which also show that 
CP is not broken and there is no first order transition. This behavior is 
also confirmed by Monte Carlo simulations \cite{bhanot}. However, the main 
interest of the results resides in their application for $N>1$ models where 
the predictions were not formerly known, although
they are compatible with some recent Monte Carlo simulations 
\cite{plekss}. In particular, they suggests that the
observed smoothing of the free energy at $\theta=\pi$ is not
entirely due to the finite volume effects but to the
finite coupling effects of the potential as has been observed in
2-dimensional QED  in the presence of an external
perturbation \cite{ae}. This effect can also be understood in 
the strong coupling expansion \cite{sei}. The cusp due to the level 
crossing of lowest energy states is removed by the   level repulsion 
generated by the potential term at first order in the strong coupling 
expansion in terms of degenerated perturbation theory \cite{ae}.

All the above arguments confirm that  CP symmetry is preserved at
$\theta=\pi$ and exclude the existence of a first order phase transition 
at $\theta=\pi$. However, the analysis does not give any clue  on the
existence or not of a second order phase transition at $\theta=\pi$. 
There are
two possibilities: 
{{i)}} Non-existence of phase transition and only a crossover
phenomena from around $\theta=\pi$ 
{{ii)}} Existence of a second order phase transition. The later effect is
what actually occurs in the case \cpa.  
One very plausible conjecture is that there is  critical N=N$_{c}$ (N$_c$=1 ?)
such that the \cp\ system  undergoes a second order phase
transition at $\theta=\pi$ for N$\leq N_{c}$ and no transition for
N$>$N$_{c}$ where there is finite mass gap. Further numerical investigations 
of this a problem would be very
interesting to clarify the behavior of the different regimes of the \cp\ 
models around $\theta=\pi$.

We thank  K. Gaw\c{e}dzki and G. Sierra for discussions.
This work was partially supported by CICyT grant AEN96-1670.

\end{document}